\documentclass[journal=jacsat,manuscript=communication]{achemso}

\usepackage[version=3]{mhchem} 
\usepackage{lineno,xcolor}

\author{Keith T. Butler}
\affiliation{Department of Chemistry, University of Bath, Claverton Down, Bath, BA2 7AY, UK}
\author{Christopher H. Hendon}
\affiliation{Department of Chemistry, University of Bath, Claverton Down, Bath, BA2 7AY, UK}
\author{Aron Walsh}
\email{a.walsh@bath.ac.uk}
\affiliation{Department of Chemistry, University of Bath, Claverton Down, Bath, BA2 7AY, UK}

\title[\texttt{achemso} demonstration]
{Electronic chemical potentials of porous metal-organic frameworks}

\begin{document}

\begin{abstract}
The binding energy of an electron in a material is a fundamental characteristic, which determines a wealth of important chemical and physical properties. For metal-organic frameworks this quantity is hitherto unknown. We present a general approach for determining the vacuum level of porous metal-organic frameworks and apply it to obtain the first ionisation energy for six prototype materials including zeolitic, covalent and ionic frameworks. This approach for valence band alignment can explain observations relating to the electrochemical, optical and electrical properties of porous frameworks.
\end{abstract}

Metal-organic frameworks (MOFs) are hybrid materials that combine both organic and inorganic functional motifs. Owing to the porous structure and large surface-area of some MOFs, they have been the subject of a concerted research effort in fields such as gas storage and catalysis.\cite{Furukawa2013,Furukawa2010,Peng2013} Recently their unique combination of optical and electronic properties has led to interest in incorporating them into photocatalytic, photovoltaic and electrochemical devices;\cite{Brozek2013,Alvaro2007,Nijem2012,Hendon2012,Dhakshinamoorthy2012,Zhan2013,Llabre2007,Horiuchi2012,Davydovskaya2012} however, the rational design of MOFs for these applications is hampered by the lack of a reference scale for the electronic levels that control these functionalities. We demonstrate a procedure, using computational chemistry, which allows us to establish values for the binding of electrons in porous frameworks, by accessing a vacuum potential level at the centre of the pores. The resulting valence band alignment for six archetype porous materials explains observations relating to the electrochemical, optical and electrical properties of these materials,\cite{Cheetham2007} and highlights a novel avenue for tuning the performance of photo- and electro-active hybrid frameworks.

The ionisation energy of an atom is well defined, \textit{i.e.} the energy required to remove an electron in the gaseous state, for example $H_{(g)} \rightarrow H^{+}_{(g)} + e^-$ ($\phi$ = 1312 kJ/mol).
For molecules, the same process occurs, but there are distinct vertical and adiabatic ionisation energies depending on whether atomic relaxation takes place. 
The ionisation energy is more difficult to define for a solid owing to the anisotropy of an electron parting the lattice. The termination of the crystal, and the associated structural and electrostatic variations, result in a large spread of measured and computed values. The bulk binding energy of an electron in a solid can be computed, for example, based on electrostatic grounds,\cite{mott-1948} or through the application of quantum chemistry with appropriate boundary conditions.\cite{scanlon-nmat}

\begin{figure}[t]
\centering
  \includegraphics[width=8.3cm]{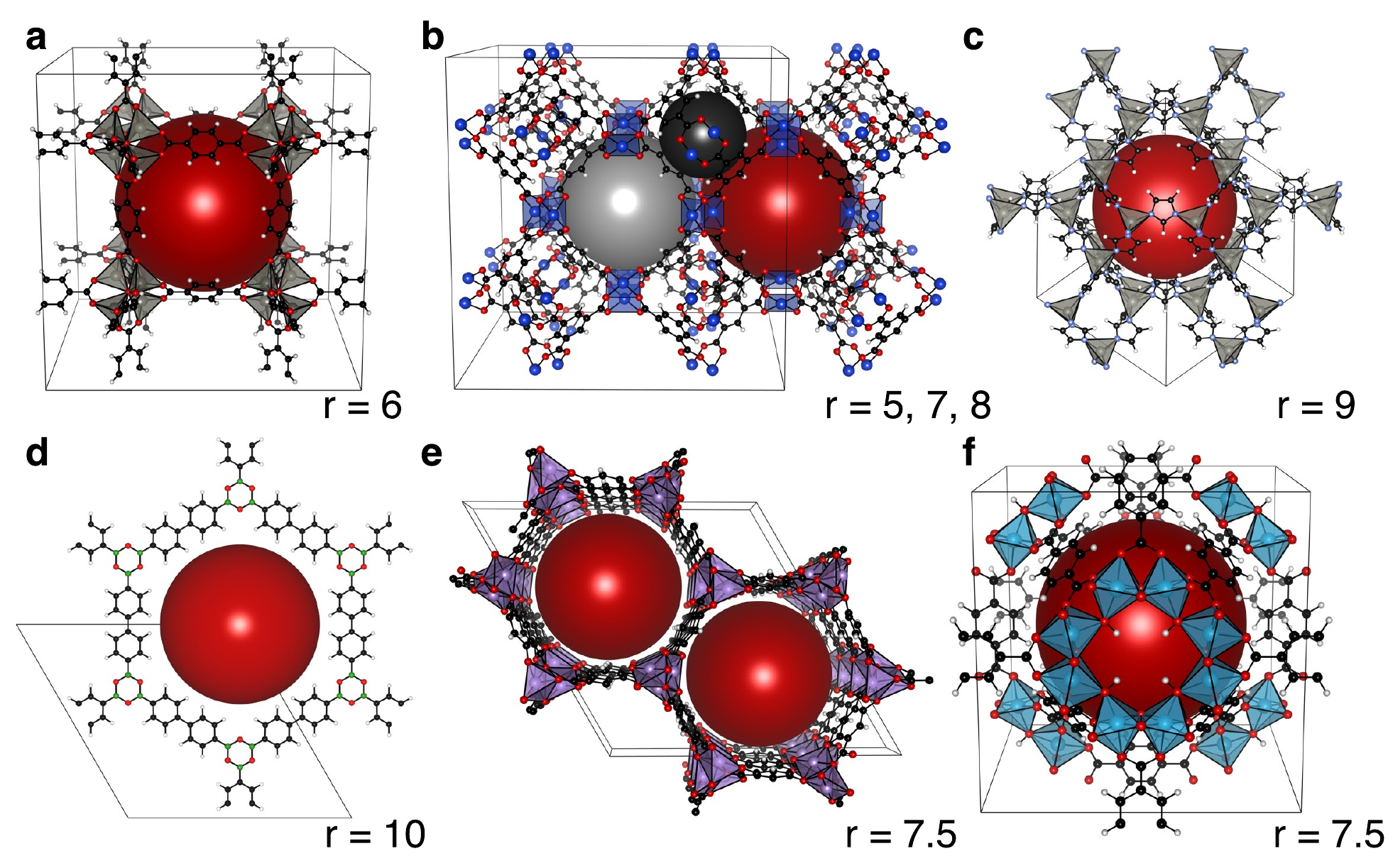}
  \caption{Structures of archetype porous frameworks. \textbf{a}: MOF-5, \textbf{b}: HKUST-1, \textbf{c}: ZIF-8, \textbf{d}: COF-1M, \textbf{e}: CPO-27-Mg and \textbf{f}: MIL-125. The largest pores of each framework are emphasised by burgundy spheres, with the pore radii (r) described in {\AA}ngstr\"{o}m.  HKUST-1 has three notably different pore sizes, emphasised with grey spheres.\cite{noteCH}}
  \label{fig1}
\end{figure}

Density functional theory (DFT) is one of the most widely used electronic structure techniques in computational materials chemistry. Indeed, the application of DFT to MOFs has resulted in hundreds of reports to date. However, the surface science of MOFs is still in its infancy; there are few models describing the atomic or electronic changes that occur at a crystal boundary. Furthermore, due to large crystallographic unit cells consisting of hundreds of atoms (see   \ref{fig1}), direct computational treatment of the surface electronic structure using quantitative methods is intractable. An alternative approach is required. 

A common feature of `designer' MOFs is porosity, with pore sizes ranging from 2 \AA~ to 50 \AA~ in radius.\cite{Deng2012} We demonstrate that the electrostatic potential at the centre of the pore provides a reference that can be used to place the electronic energy levels of MOFs on a common energy scale. Following the validation of this approach, we report the valence band energy of six familiar frameworks (\ref{fig1}), including a key metal-organic framework (MOF-5)\cite{Tranchemontagne2008}, two of the highest performing gas-storage coordination frameworks (CPO-27-Mg and HKUST-1),\cite{Dietzel2010,Chui1999} a covalent organic framework (COF-1M),\cite{Cote2005,Lukose2010} a zeolitic imidazolate framework (ZIF-8)\cite{Park2006} and a material of Institut Lavoisier (MIL-125)\cite{Dan-Hardi2009} that differ in both local and extended connectivity. This approach defines the reference potentials necessary for rational design of MOFs for electronic devices and photocatalytic applications. The generality and low computational overhead make it suitable for incorporation into materials screening procedures.\cite{Wilmer2011}

\begin{figure}[hb!]
\centering
  \includegraphics[width=8.3cm]{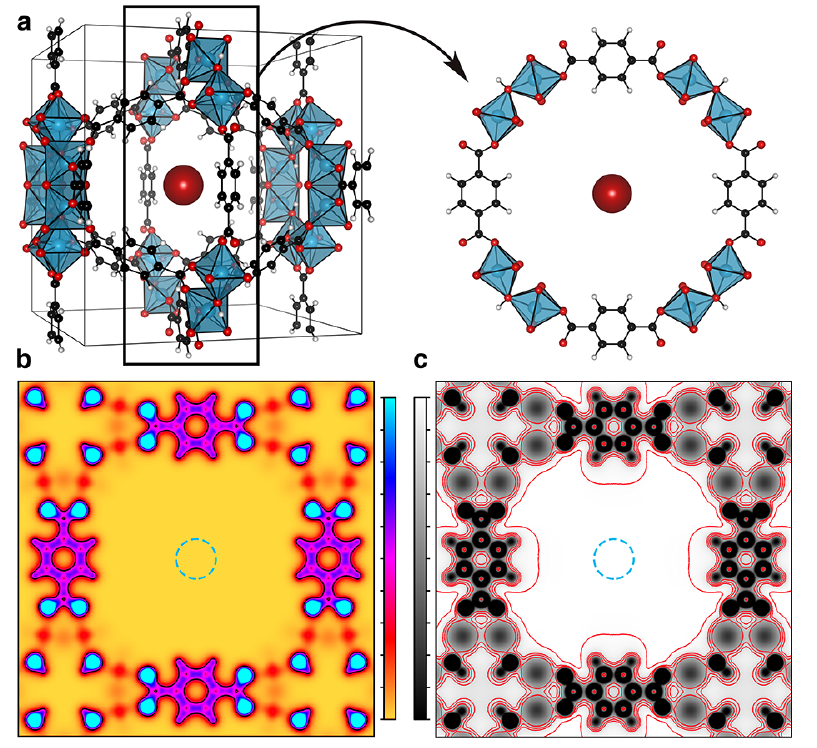}
  \caption{Illustration of the procedure used to calculate an electrostatic reference potential for metal-organic frameworks. (a) The structure of MIL-125, with the red spherical electrostatic probe, $r$ = 2 \AA, shown at the pore centre. (b) (001) slice through the valence electron density of MIL-125, drawn from yellow (0 e/\AA$^3$) to blue (0.5 e/\AA$^3$). (c) (001) slice through the total electrostatic potential of MIL-125, drawn from black (-29.45 eV) to white (2.45 eV), with respect to the pore centre. Red contours are shown from 2.45 eV to -10 eV in 1 eV intervals. The probed region is shown with a blue dashed circle in (b) and (c).}
  \label{fig2}
\end{figure}

Chemical interactions are predominately ``near-sighted".\cite{Prodan2005} The valence electron density taken through a plane of MIL-125, which is composed of TiO$_2$ octahedra and 1,4-benzenedicarboxylate (\textbf{bdc}), is shown in   \ref{fig2}. The electrons are confined to the hybrid framework, with strong localisation at anionic (oxide) centres. The associated electrostatic potential resulting from the nuclear and electronic distributions is also shown. These interactions extend further from the atomic centres, \textit{e.g.} the potential energy of interacting quadrupole moments varies with distance as $r^{-5}$; however, they rapidly decay towards the centre of the pore and the electrostatic potential plateaus to a constant. The same behaviour is observed in all six frameworks studied.  

To ensure a robust reference energy, we compute the spherical average of the electrostatic potential at the pore centre:
\begin{equation}
\label{eqn1} \Phi_{av}(r)=\frac{1}{V}\int\limits_V\Phi(r')d^3r'
\end{equation}
The mean and variance of the potential values within the sphere are used to assess the convergence, and furthermore compute the principal components of the electric field tensor ($\textbf{E}_{xx}, \textbf{E}_{yy}, \textbf{E}_{zz}$). A radius of 2 \AA ~ is used, and the results presented are insensitive to this choice up to 4 \AA. For the six cases studied, the variance is within $1\times10^{-4}$ V, while the electric field falls within $1\times10^{-4}$ V/\AA. The full data set including electrostatic potential plots for each framework and details for obtaining the analysis code are included as Supplementary Information (S1).

In the absence of strong long-range electric fields, the plateau in the electrostatic potential represents a sound approximation to the vacuum level; however, there may be special cases where it represents a local level influenced by the polarity of the terminal groups around the pore. This distinction is analogous to the difference between the bulk binding energy of an electron in a crystal and the anisotropic ionisation potential associated with a particular surface termination. 

\begin{figure*}[t!]
\centering
  \includegraphics[width=18cm]{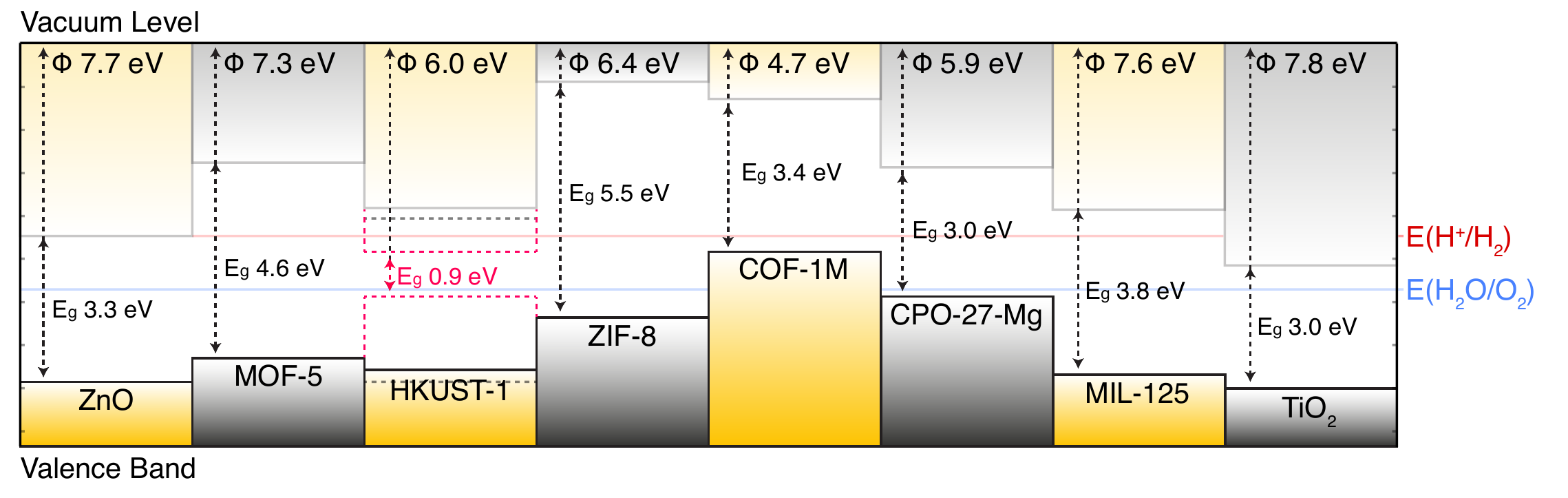}
  \caption{Predicted vertical ionisation energy of six prototype porous metal-organic frameworks with respect to a common vacuum level (determined by the value of the electrostatic potential at the centre of an internal pore). Note that for HKUST-1 values are shown for the ground state antiferromagnetic singlet (solid lines), triplet state (black dashed lines) and the closed shell singlet (pink dotted lines). The values were calculated using density functional theory, employing a hybrid exchange-correlation functional (HSE06), and with periodic boundary conditions used to represent the perfect solid. The redox potentials of water are drawn as horizontal lines, and values for the inorganic solids (wurtzite) ZnO and (rutile) TiO$_2$ are taken from recent embedded-cluster calculations.\cite{Catlow2008,scanlon-nmat}}
  \label{fig3}
\end{figure*}

The alignment of the six frameworks, following the procedure outlined above, is shown in   \ref{fig3}. 
The valence band energies are between 7.64 eV (MIL-125) and 4.67 eV (COF-1M), which fall within the range expected for solid-state materials.

HKUST-1 represents the most challenging framework considered since it contains three distinct pore topologies ranging from 5 $-$ 8 \AA ~ in radius (\ref{fig1}). The vacuum potential is converged within 0.1 V for the smallest pore and within 0.01 V for the largest.  Note that there are three possible spin configurations arising from the Cu(II) 3\textit{d}$^9$ states at the top of the valence band; in increasing energy, the open-shell singlet (antiferromagnetic state), the triplet (ferromagnetic state) and the closed shell singlet (a Cu-Cu $\delta$ bond).  Depending on the method of IP measurement, different values can be obtained.  For instance, in a recent study by Lee and co-workers, an IP of 5.43 eV was measured using cyclic voltametry (CV) of an iodine-doped film.\cite{Lee2013,Loera-Serna2012}  Cyclic voltammetry probes the redox processes in solution: the energy of the highest accessible configuration (pink dotted line, Cu-Cu $\delta$  bond, \ref{fig3}) will be probed. It should be noted that CV measurements are highly sensitive to surface and interface effects that our method implicitly avoids; nonetheless, the agreement is satisfying. Measurements of the IP using ultraviolet photoelectron spectroscopy (UPS) would be more comparable to our predictions (\textit{i.e.} black solid line, \ref{fig3}).

Our calculated HKUST-1 IP explains the recently reported increase in electroactivity by the inclusion of tetracyanoquinodimethane (\textbf{TCNQ}), which bridges adjacent Cu-Cu motifs in the largest pore (r $=$ 8 \AA).  Talin and co-workers report the molecular IP of \textbf{TCNQ} at 7.7 eV, this value coincides with our solid state IP for the HKUST-1 host framework (7.6 eV for the antiferromagnetic state).\cite{Talin2013}  The result is an `Ohmic contact', establishing a direct channel for electronic flow throughout the framework.  Based on this result one can predict optimal band offsets between MOFs and guest molecules, allowing rational design for a host of applications (\textit{e.g.} catalysis, optoelectronics etc.).

The MIL-125 framework contains cyclic octamers of TiO$_2$ octahedra. Previous analysis has shown that the valence band is dictated by the \textbf{bdc} ligand, while the conduction band is formed of empty Ti \textit{d} and O \textit{p} orbitals.  The valence band of the binary metal oxide TiO$_2$ has been placed at 7.8 - 8.3 eV below the vacuum level, depending on the polymorph.\cite{scanlon-nmat} The predicted value of 7.64 eV for MIL-125 can be explained by the lower binding energy of the aromatic $\pi$ system. The larger band gap of MIL-125 places its conduction band above that of TiO$_2$, which can be understood from the reduced dimensionality (quantum confinement) of the Ti sub-lattice. 
MIL-125 has electronic potentials suitable for application as a photocatalyst, with the an electron affinity that is lower than the water reduction potential. Engineering of the valence band energy through ligand functionalisation has recently been demonstrated,\cite{Hendon2013JACS} which could be used to produce a hybrid photocatalyst active in the visible range of the electromagnetic spectrum. 

MOF-5 is composed of the same \textbf{bdc} linkers as MIL-125, but the inorganic building blocks are replaced by tetrahedra of ZnO. As the valence band is controlled by \textbf{bdc}, the ionisation energy of 7.30 eV is close to that of MIL-125, despite their distinct crystal structures. This value is again lower than the parent inorganic oxide; the valence band of ZnO has been placed at 7.71 eV below vacuum.\cite{Catlow2008} Due to the larger band gap of MOF-5, again from confinement of the ZnO sub-lattice, the electron affinity is lower than the parent oxide and well above the water-reduction potential. 

The three other frameworks display distinctly smaller ionisation energies of 4.67 eV (COF-1M), 5.87 eV (CPO-27-Mg) and 6.37 eV (ZIF-8).
Similar to MOF-5, the structure of ZIF-8 contains tetrahedra of Zn, but with the O anions replaced by N. Owing to the lower binding energy of the N 2\textit{p} orbitals, which form the imidizole linker, the valence band energy is significantly higher than both MIL-25 and MOF-5.
CPO-27-Mg contains a linking unit similar to MIL-125, 2,4-dihydroxy-\textbf{bdc}, but it is an electron-rich analogue to \textbf{bdc} that further reduces the ionisation energy.  CPO-27-Mg has potential for electronic activation through guest molecule inclusion, some interesting candidates based on an IP matching argument have been recently reported by Hendon \textit{et al.}\cite{Hendon2013CS}
COF-1M, a biphenyl hypothetical analogue of COF-1, has the highest valence band of the examined MOFs.  The biphenyl and boroxine units produce extended $\pi$ conjugation that gives rise to p-type hole-mediated conductivity.

Knowledge of the electronic chemical potentials has impact beyond the individual electron and removal energies. 
The design and optimisation of novel semiconductors has rapidly progressed through doping limit rules based on the energy of the valence and conduction bands,\cite{PhysRevLett.84.1232} e.g. a high valence band (low ionisation potential) should result in effective p-type behaviour. Our results for COF-1M demonstrate that these rules are also applicable to organic frameworks. 
Concepts such as universal alignment of defect levels\cite{VandeWalle2003} can now be applied to a new class of materials and MOFs may be selected or designed to provide Ohmic or Schottky contacts in electrical devices. 
An intriguing observation from the computed alignments for MOF-5 and MIL-125 is that the band offsets with their parent inorganic compounds are of Type-II,\cite{yu-05} so that a oxide/MOF heterojunction could be exploited to separate electron and hole carriers for application in photoconvertors.

In summary, an approach has been developed to place the electronic states of porous metal-organic frameworks on a common energy scale, based upon quantities obtained from electronic structure calculations. The method can be integrated into high-throughput workflows. We report the electron removal energies for six archetypal metal-organic frameworks, explaining the physical origin of conductivity (COF-1M) and photocatalyic behaviour (MIL-125). Knowledge of the electronic chemical potentials provides a roadmap for designing high-performance electro-active metal-organic frameworks. 

\section{Methods}
\subsection{Electronic structure}
\textit{Computational Methods}: All electronic and structural calculations were performed within the Kohn-Sham density functional theory (DFT) framework. Born-von K{\'a}rm{\'a}n boundary conditions were employed to represent a framework infinitely repeating in each direction, with no surface termination. The Vienna \textit{ab initio} simulation package (\textit{VASP}),\cite{Kresse1996} a plane-wave basis set code (with PAW scalar-relativistic pseudopotentials), was employed for crystal and electronic structure optimisation. $\Gamma$-point sampling of the Brillouin zone was used for each of the frameworks, which is sufficient considering their large real-space dimensions. A 500 eV plane-wave cut-off was found to be suitable for convergence of electronic wavefunctions to give total energies within 0.01 eV/atom. Starting with the experimentally determined unit cells of the frameworks, both lattice parameters and atomic positions were relaxed with the semi-local Perdew-Burke-Ernzerhof exchange-correlation functional revised for solids (PBEsol).\cite{Perdew2008} The resulting structures were found to be within 1\% of the experimental values. 

The key electronic properties, including electron density, electrostatic potential and band gap, were computed using a hybrid exchange-correlation functional (HSE06)\cite{Heyd2003,Krukau2006} with 25\% of the short-range semi-local exchange replaced by the exact non-local Hartree-Fock exchange. 

\subsection{Electrostatic alignment}
In contrast to molecular quantum-chemical calculations, within periodic boundary conditions, the electronic eigenvalues resulting from the solution of the Kohn-Sham equations are given with respect to an internal reference (for \textit{VASP} it is the average electrostatic potential of the repeating cell). The consequence is that absolute values of band energies cannot be compared between two or more frameworks: there is no common vacuum level. 
It should be noted that for solids, unlike finite systems, the highest occupied Kohn-Sham eigenvalue and the electron removal energy ($N \rightarrow N-1$ system) are equivalent in the dilute limit. 

For the reference electrostatic potential we use a spherical average of the Hartree potential in a sphere of radius 2 \AA ~ with an origin at the centre of the MOF pore.  The analysis code for this calculation, which can also calculate planar and macroscopic averages of electrostatic potentials and charge densities, is freely available.\cite{KTB}  The electrostatic potential was sampled on a grid of mesh density $>$ 14 points/\AA. Further details of the approach are provided as Supplementary Information. 

\acknowledgement
We acknowledge useful discussions on this topic with C. Mellot-Draznieks, M. D. Allendorf and D. Tiana. KB, CHH and AW are supported by the EPSRC (Grant No. EP/J017361/1), the ERC (Grant No. 277757) and the Royal Society. The work benefited from the University of Bath's High Performance Computing Facility, and access to the HECToR supercomputer through membership of the UKs HPC Materials Chemistry Consortium, which is funded by EPSRC (Grant No. EP/F067496).

\suppinfo
Details of the electrostatic alignment procedure and equilibrium crystal structures are available free of charge via the Internet at http://pubs.acs.org.

\bibliography{MOF_WF}

\begin{mcitethebibliography}{41}
\bibitem[not()]{noteCH}
MIL-125 and MOF-5 have two unqiue pores of similar diameter. In this study, we
  sampled the largest; however, sampling the smaller pores of these two
  materials yields the same predicted IPs.\relax
\mciteBstWouldAddEndPunctfalse
\mciteSetBstMidEndSepPunct{\mcitedefaultmidpunct}
{}{\mcitedefaultseppunct}\relax
\EndOfBibitem
\end{mcitethebibliography}


\end{document}